# A NEW VARIABLE STEP-SIZE ZERO-POINT ATTRACTING PROJECTION ALGORITHM*

*Jianming Liu and Steven L Grant*[1]

Department of Electrical and Computer Engineering, Missouri University of Science and Technology, Rolla, Missouri 65409.

## ABSTRACT

This paper proposes a new variable step-size (VSS) scheme for the recently introduced zero-point attracting projection (ZAP) algorithm. The proposed variable step-size ZAPs are based on the gradient of the estimated filter coefficients' sparseness that is approximated by the difference between the sparseness measure of current filter coefficients and an averaged sparseness measure. Simulation results demonstrate that the proposed approach provides both faster convergence rate and better tracking ability than previous ones.

## 1. INTRODUCTION

In many practical applications, such as the network echo cancellation, the impulse response is usually sparse, which means only a small percentage of coefficients are active and most of the others are zero or close to zero [1]. Classical normalized least-mean-square (NLMS) suffers from slow convergence rate and many adaptive algorithms have been proposed to exploit the sparse nature of the system to improve performance. These include the proportionate family, in which the most popular proportionate adaptive algorithms are proportionate NLMS (PNLMS) [2], improved proportionate NLMS (IPNLMS) [3] and mu-law proportionate NLMS (MPNLMS) [4], etc.

Recently, a new LMS algorithm with $l_0$ norm constraint was proposed to accelerate sparse system identification [5]. It applied the constraint to the standard LMS cost function and when the solution is sparse, the gradient descent recursion will accelerate the convergence of near-zero coefficients of the sparse system. Another similar approach was proposed in [6], but it is based on $l_1$ norm penalty.

The above scheme was referred as *zero-point attraction projection* (ZAP) in [7] and their performance analysis have been report in [8]-[10]. Analysis showed that the step-size of the ZAP term denotes the importance or the intensity of attraction. A large step-size for ZAP results in a faster convergence, but the steady-state misalignment also increases with a large step-size. So, the step-size of ZAP is also a trade-off between convergence rate and steady-state misalignment, which is similar to the step-size trade-off of LMS. However, the variable step-size (VSS) ZAP algorithms have not been exploited too much and most of the previous algorithms are based on theoretical results which could not be calculated in practice [9]-[10].

As far as we know, the only variable step-size scheme for ZAP was proposed by You, etc. in [11], in which it was initialized to be a large value and reduced by a factor when the algorithm is convergent. However, this heuristic strategy cannot track the change in the system response due to the very small steady-state step-size.

This paper is organized as follows. Section 2 reviews the recently proposed ZAP and VSS algorithm for ZAP, and in Section 3 we present the proposed VSS ZAP algorithm. The simulation results and comparison to the previous algorithms are presented in Section 4. Finally conclusions are drawn in Section 5.

## 2. REVIEW OF VSS ZAP

In the scenario of echo cancellation, the far-end signal $\boldsymbol{x}(n)$ is filtered through the room impulse response $\boldsymbol{h}(n)$ to get the echo signal $y(n)$.

$$y(n) = \boldsymbol{x}(n) * \boldsymbol{h}(n) = \boldsymbol{x}_n^T \boldsymbol{h}_n \quad (1)$$

where

$\boldsymbol{x}_n = [x(n)x(n-1)\cdots x(n-L+1)]^T$, $\boldsymbol{h}_n = [h_0 h_1 \cdots h_{L-1}]^T$,

and $L$ is the length of echo path. This echo signal is added to the near-end signal $v(n)$ (including both speech and back ground noise, etc.) to get the microphone signal $d(n)$,

$$\begin{aligned} d(n) &= \boldsymbol{x}(n) * \boldsymbol{h}(n) + v(n) \\ &= \boldsymbol{x}_n^T \boldsymbol{h}_n + v(n). \end{aligned} \quad (2)$$

*This work was performed under the Wilkens Missouri Endowment.

[1] Formerly Steven L Gay.

We define the estimation error of the adaptive filter output with respect to the desired signal as

$$e(n) = d(n) - \mathbf{x}_n^T \mathbf{w}_n \quad (3)$$

This error, $e(n)$ is used to adapt the adaptive filter $\mathbf{w}(n)$. The LMS algorithm updates the filter coefficients as below [1]:

$$\mathbf{w}(n) = \mathbf{w}(n-1) + \mu \mathbf{x}_n^T e(n) \quad (4)$$

in which $\mu$ is the step-size of adaption. The LMS algorithm with $l_0$ norm constraint added a *zero attractor* and update is as below [5]:

$$\begin{aligned} \mathbf{w}(n) &= \mathbf{w}(n-1) + \mu \mathbf{x}_n^T e(n) \\ &\quad - \kappa\beta \,\mathrm{sgn}(\mathbf{w}(n-1)) \otimes e^{-\beta|\mathbf{w}(n-1)|} \end{aligned} \quad (5)$$

where $\kappa$ is the step-size of zero attractor, $\beta$ is a constant, and $\otimes$ means component-wise multiplication. $\mathrm{sgn}(\cdot)$ is a component-wise sign function defined as

$$\mathrm{sgn}(x) = \begin{cases} \frac{x}{|x|}, & x \neq 0; \\ 0, & elsewhere. \end{cases} \quad (6)$$

The LMS algorithm with $l_1$ norm constraint was proposed in [6], and its update equation is

$$\mathbf{w}(n) = \mathbf{w}(n-1) + \mu \mathbf{x}_n^T e(n) - \kappa\,\mathrm{sgn}(\mathbf{w}(n-1)) \quad (7)$$

The variable step-size used in [11] is rather direct: $\kappa$ is initialized to be a large value, and reduced by a factor $\eta$ when the algorithm is convergent. This reduction is conducted until is sufficiently small, i.e. $\kappa < \kappa_{\min}$, which means that the error reaches a low level. However, as mentioned in the introduction, this heuristic strategy will not react to a change in the system response since it will get stuck due to the very small steady-state step-size. Therefore, in order to solve this issue, we will propose a variable step-size ZAP algorithm in next section which could both converge fast and track the change efficiently.

## 3. PROPOSED VSS ZAP

Our proposed new variable step-size ZAP algorithm is based on the measurement of the sparseness gradient approximated by the difference between the sparseness measure of current filter coefficients and an averaged sparseness measurement. Therefore, the proposed VSS ZAP can track the change of system quickly and demonstrate a good balance between fast convergence rate and lower stable state misalignment.

For the measurement of sparsity, we could use a class of sparsity-inducing penalties. The penalty is defined as

$$J(\mathbf{w}(n)) = \sum_{i=1}^{L} G(w_i(n)) \quad (8)$$

where $G(\cdot)$ belongs to a class of sparseness measures [12]. Some commonly used sparseness measures are introduced in Table 1, where $\chi_P$ denotes the indicator function:

$$\chi_P = \begin{cases} 1 & P \text{ is true;} \\ 0 & P \text{ is false.} \end{cases} \quad (9)$$

**Table 1.** Sparseness Measures in [12]

| No. | $G(t)$ | Param. Require. |
|---|---|---|
| 1. | $\|t\|$ | ------ |
| 2. | $\frac{\|t\|}{(\|t\|+\sigma)^{1-p}}$ | $0 \le p < 1$ |
| 3. | $1-e^{-\sigma\|t\|}$ | $\sigma > 0$ |
| 4. | $\ln(1+\sigma\|t\|)$ | $\sigma > 0$ |
| 5. | $\mathrm{atan}(\sigma\|t\|)$ | $\sigma > 0$ |
| 6. | $(2\sigma\|t\|-\sigma^2 t^2)\chi_{\|t\|\le\frac{1}{\sigma}} + \chi_{\|t\|>\frac{1}{\sigma}}$ | $\sigma > 0$ |

They are mainly from [12], but they are still included in this paper for completeness. Besides to the sparseness measures listed in Table. 1, another popular measurement of channel sparsity was proposed in [13] as below. For a channel $\mathbf{h}(n)$, its sparsity $\varepsilon(\mathbf{h}(n))$ can be defined as

$$\varepsilon(\mathbf{h}(n)) = \frac{L}{L-\sqrt{L}}\left(1 - \frac{\|\mathbf{h}(n)\|_1}{\sqrt{L}\|\mathbf{h}(n)\|_2}\right). \quad (10)$$

where $L > 1$ is the length of the channel $\mathbf{h}(n)$, and $\|\mathbf{h}(n)\|_1$ and $\|\mathbf{h}(n)\|_2$ are the $l_1$ norm and $l_2$ norm of $\mathbf{h}(n)$.

The value of $\varepsilon(\mathbf{h}(n))$ is between 0 and 1. For a sparse channel the value of sparsity is close to 1 and for a dispersive channel, this value is close to 0. Therefore, this property could be used to remove the ZAP term when the channel is dispersive, which is preferable. Instead of calculating the sparseness of the real channel, the sparsity of the current adaptive filter $\mathbf{w}(n)$ is estimated as [13].

$$\varepsilon(\mathbf{w}(n)) = \frac{L}{L-\sqrt{L}}\left(1 - \frac{\|\mathbf{w}(n)\|_1}{\sqrt{L}\|\mathbf{w}(n)\|_2}\right). \quad (11)$$

The gradient of sparseness measure could be approximated by the difference between the sparseness measure of current filter coefficients and an averaged sparseness measurement. The averaged sparseness measure could be estimated adaptively with a forgetting factor as below:

$$\phi(n)=(1-\lambda)\phi(n-1)+\lambda J(\boldsymbol{w}(n)),\ 0<\lambda<1 \qquad (12)$$

The difference between the sparseness measure of current filter coefficients and the averaged sparseness measurement is calculated by:

$$\delta(n)=J(\boldsymbol{w}(n))-\phi(n-1) \qquad (13)$$

Similar to [14], in order to obtain a good and stable estimate of the gradient, a long-term average using infinite impulse response filters is used to calculate the proposed variable step-size as below:

$$\kappa(n)=(1-\alpha)\kappa(n-1)+\alpha\gamma|\delta(n)|,\ 0<\alpha<1 \qquad (14)$$

in which $\alpha$ is a smoothing factor and $\gamma$ is a correction factor.

## 4. SIMULATION RESULTS

In this section, we do the results of computer simulations in the scenario of echo cancellation. In order to evaluate the performance of our proposed VSS ZAP in both sparse and dispersive impulse response, we use a sparse impulse response as in Fig. 2 and a dispersive random impulse response as in Fig. 3. They are both with the same length, *L=512*, and the LMS adaptive filter is the same length.

The convergence state of adaptive filter is evaluated using the normalized misalignment which is defined as

$$10\log_{10}(\|\boldsymbol{h}-\boldsymbol{w}\|_2/\|\boldsymbol{h}\|_2)\,. \qquad (15)$$

In this simulation, we compare the proposed VSS algorithm to LMS, LMS with fixed step-size ZAP and You's VSS ZAP in [11]. For the $l_1$ norm constraint ZAP, we will use the No. 1 sparseness measure in Table 1 for simple, and in order to save computation efforts, for the $l_0$ norm constraint ZAP, we will use the same No. 3 sparseness measure as in Table 1. Meanwhile, to evaluate the performance under dispersive system, we also use the measurement of sparsity as in (11), and compare it to the above algorithms.

The input is white Gaussian noise signal and independent white Gaussian noise is added to the system background with a signal-to-noise ratio, SNR = 30 dB. The parameters of VSS ZAPs are chosen to allow all the VSS ZAPs to have similar final steady-state misalignment (about -25 dB) as standard LMS.

In order to compare the tracking, we simulate the echo path change at sample 5000 by switching to another sparse impulse response. We plot the normalized misalignment and variable step-size for $l_1$ norm constraint ZAP as in Fig. 4a and Fig. 4b.

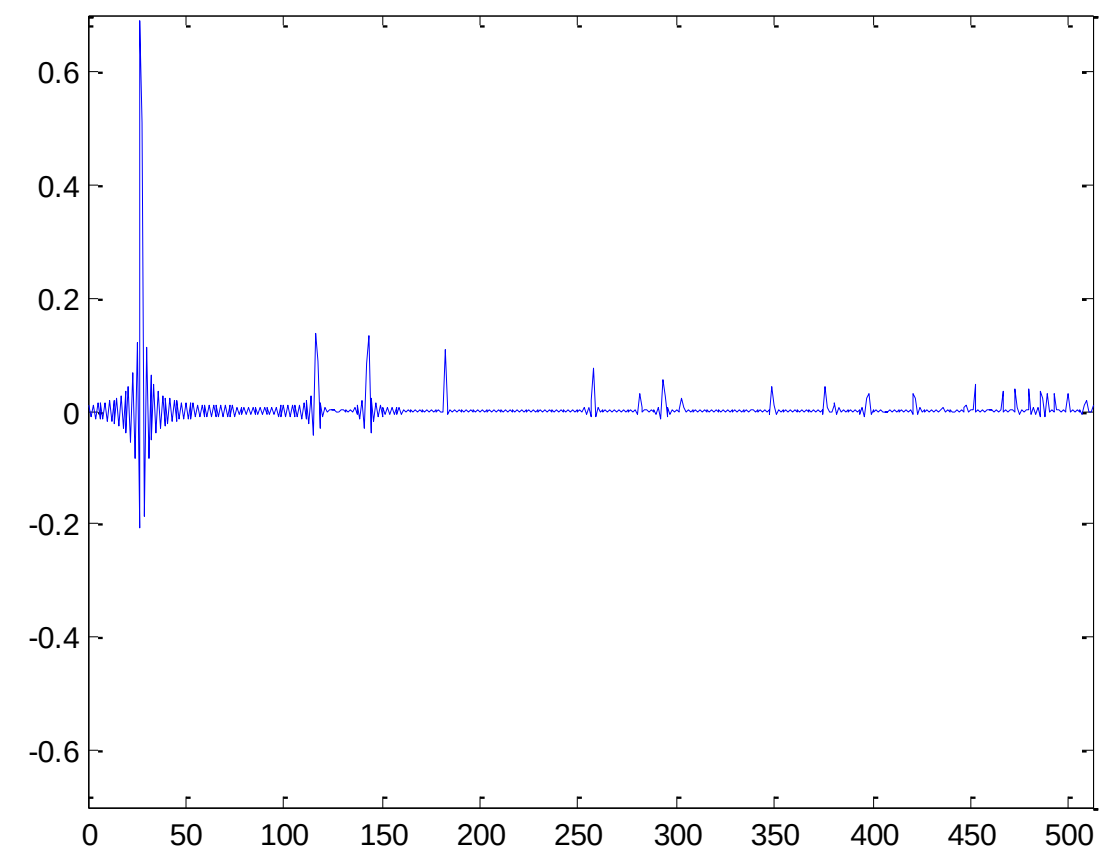

**Fig. 2** Sparse impulse response

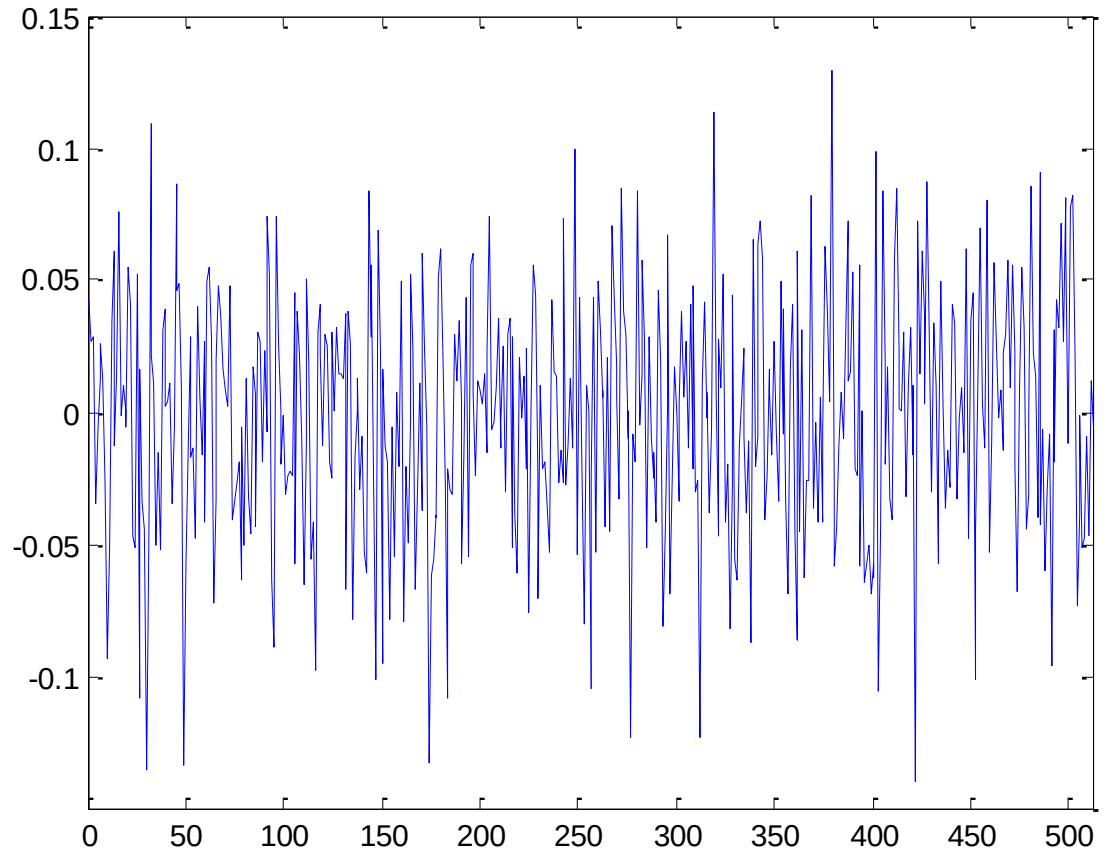

**Fig. 3** Dispersive random impulse response.

Similarly, the normalized misalignment and variable step-size for $l_0$ norm constraint ZAP are plotted in Fig. 5a, and Fig. 5b. It should be noted that we call the sparseness measure from Table. 1 as proposed VSS 1, and the measurement of sparsity in (11) as proposed VSS 2. We could clearly observe that the proposed VSS ZAPs are superior to standard LMS, fixed step-size ZAP LMS and previous You's VSS ZAP in the terms of convergence rate, and the tracking ability.

Finally, in order to demonstrate the performance for dispersive channel, we switch the sparse echo path in Fig. 2 to a dispersive random echo path as in Fig. 3. The performance and VSS for $l_1$ norm constraint ZAP are plotted in Fig. 6a and Fig. 6b, and $l_0$ norm constraint ZAP in Fig. 7a and Fig. 7b. It is clear that the sparsity measurement in (11) could remove the impact of ZAP term under non-sparse system and performs better than the sparseness measure in Table 1. This is because the steady-state step-size of proposed VSS 1 ZAP is bigger which will cause performance degradation under non-sparse system.

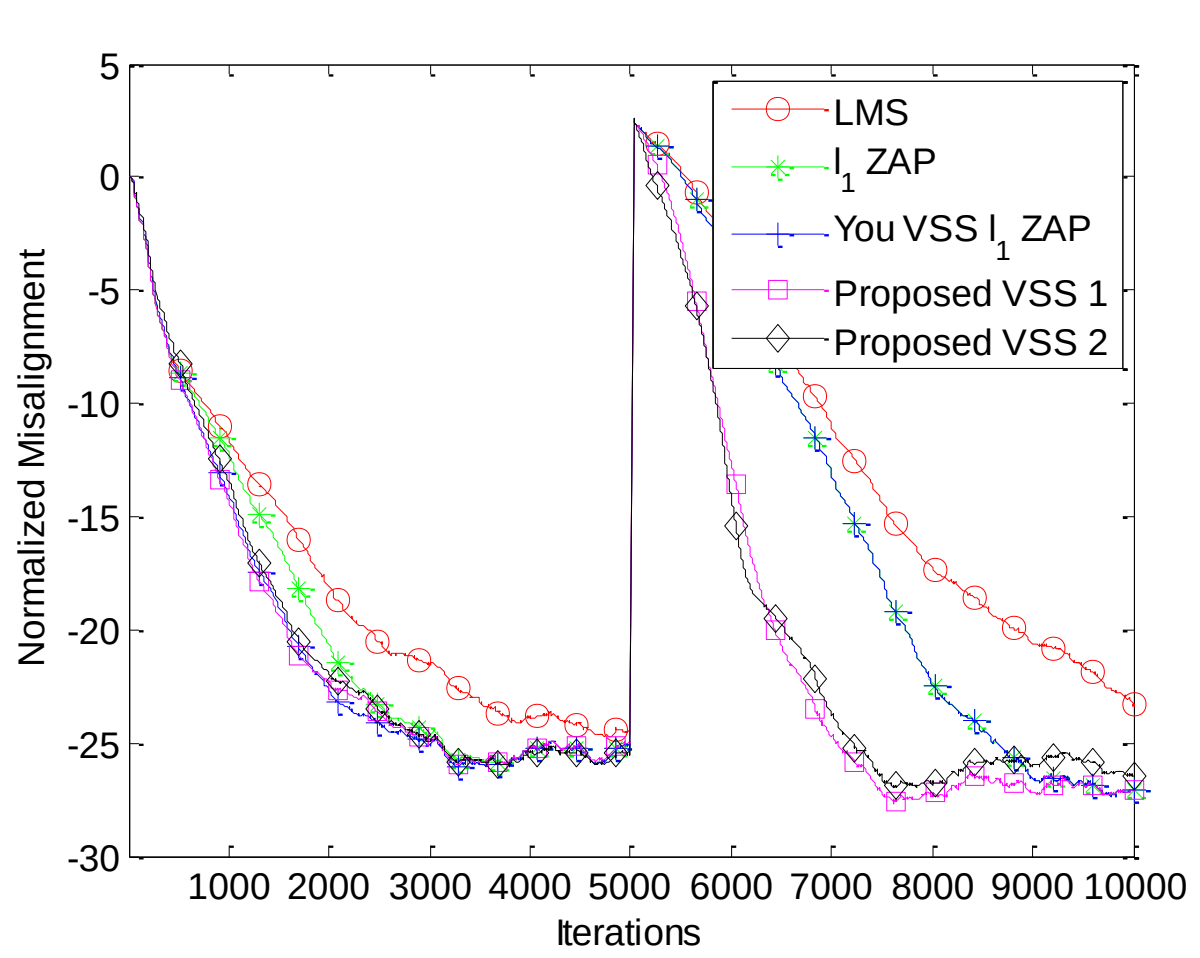

**Fig.4a** Comparison of normalized misalignment for $l_1$ norm constraint ZAP under sparse system.

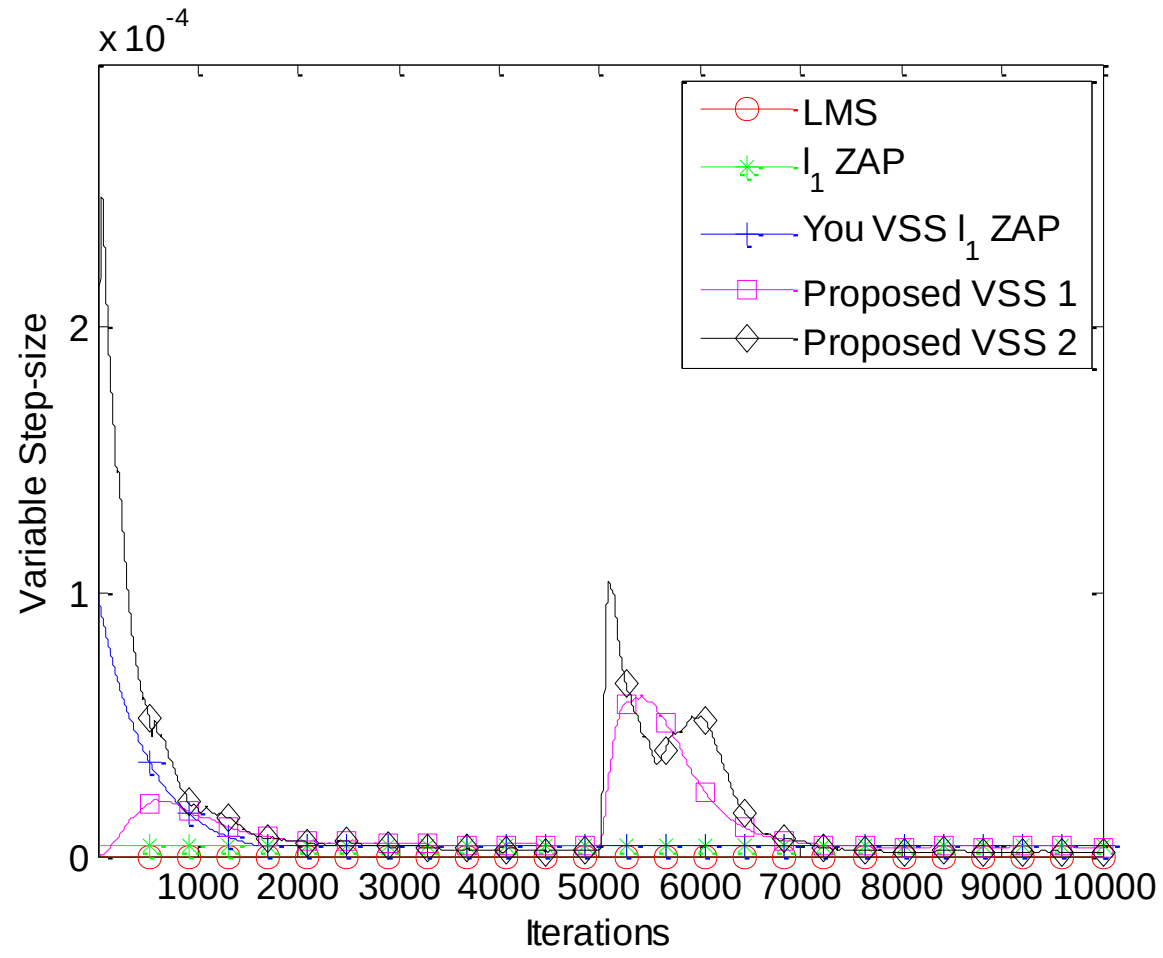

**Fig.4b** Comparison of variable step-size for $l_1$ norm constraint ZAP under sparse system.

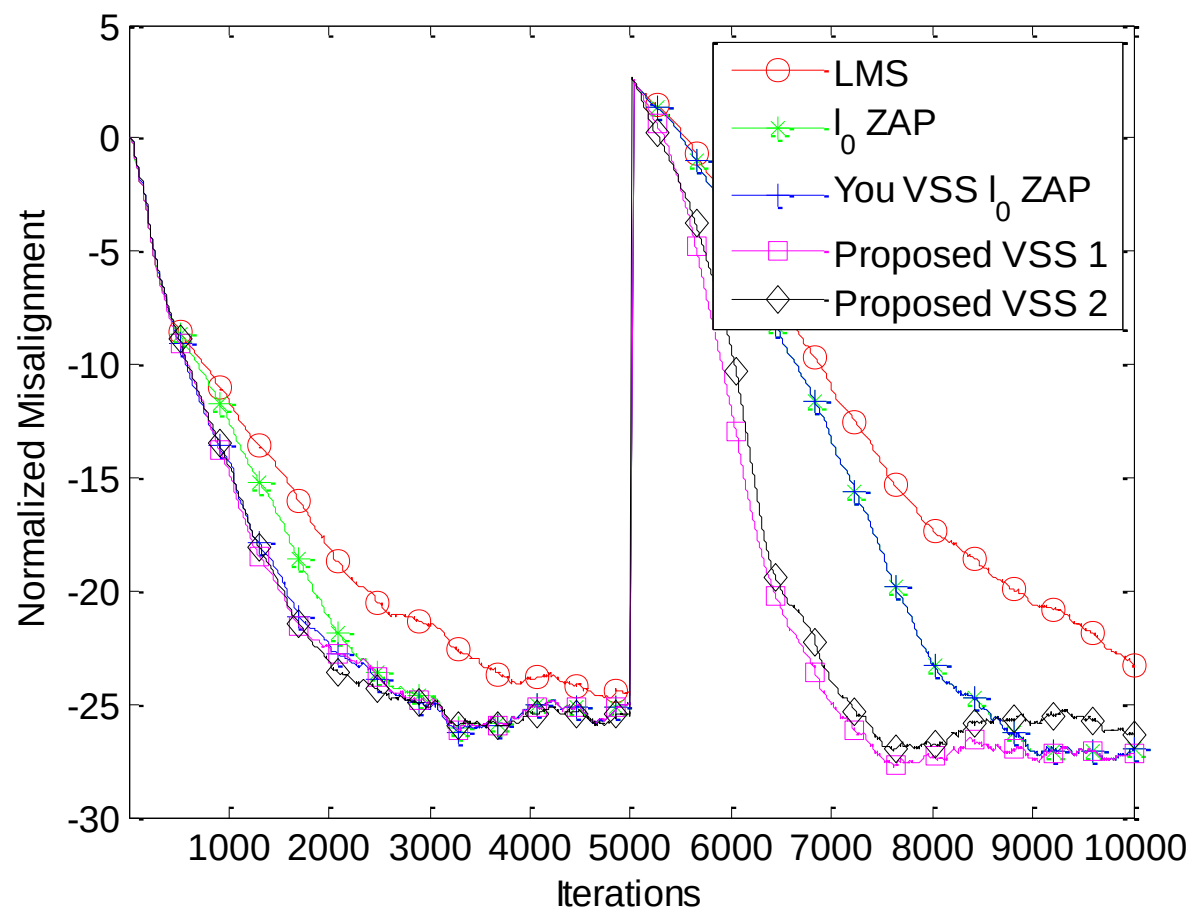

**Fig.5a** Comparison of normalized misalignment for $l_0$ norm constraint ZAP under sparse system.

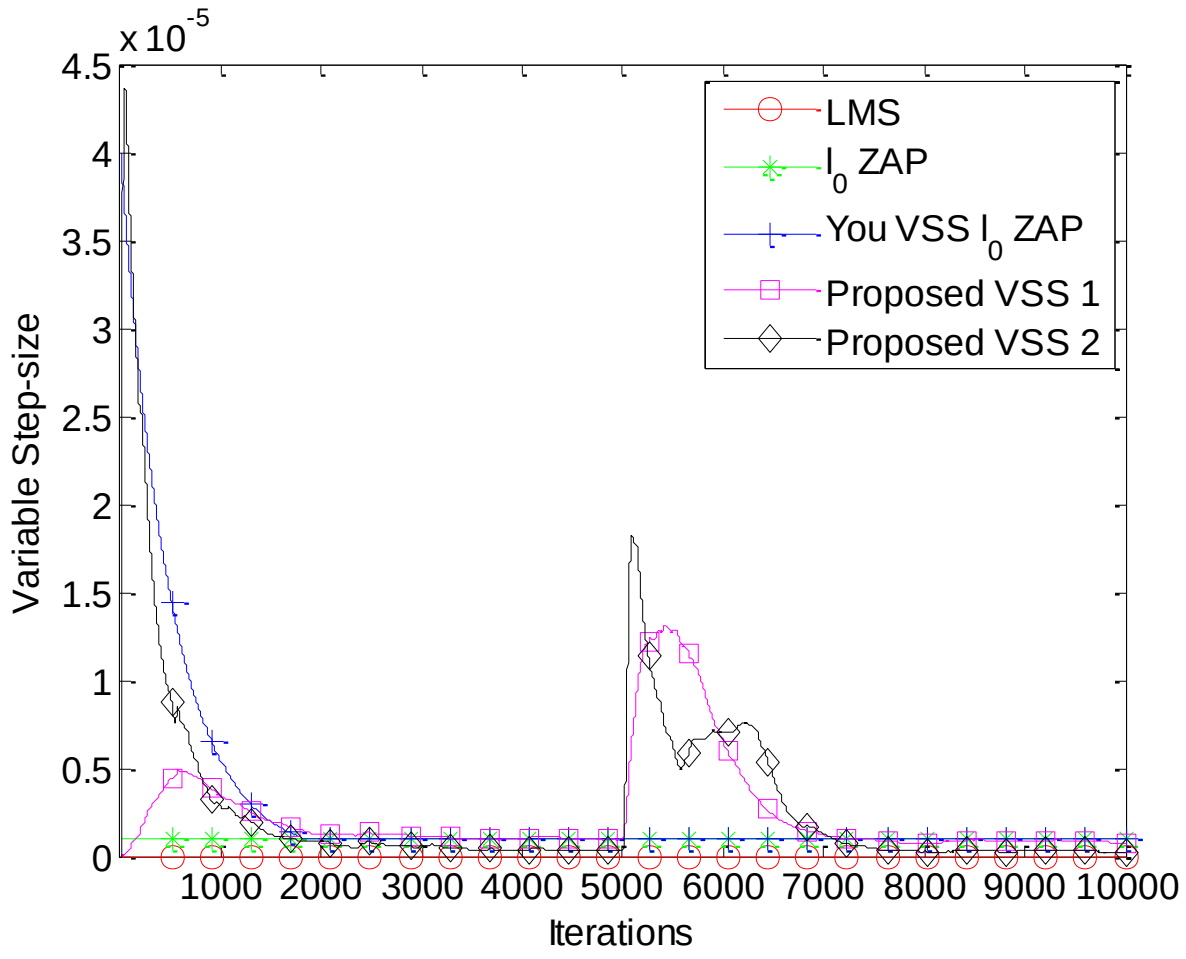

**Fig.5b** Comparison of variable step-size for $l_0$ norm constraint ZAP under sparse system.

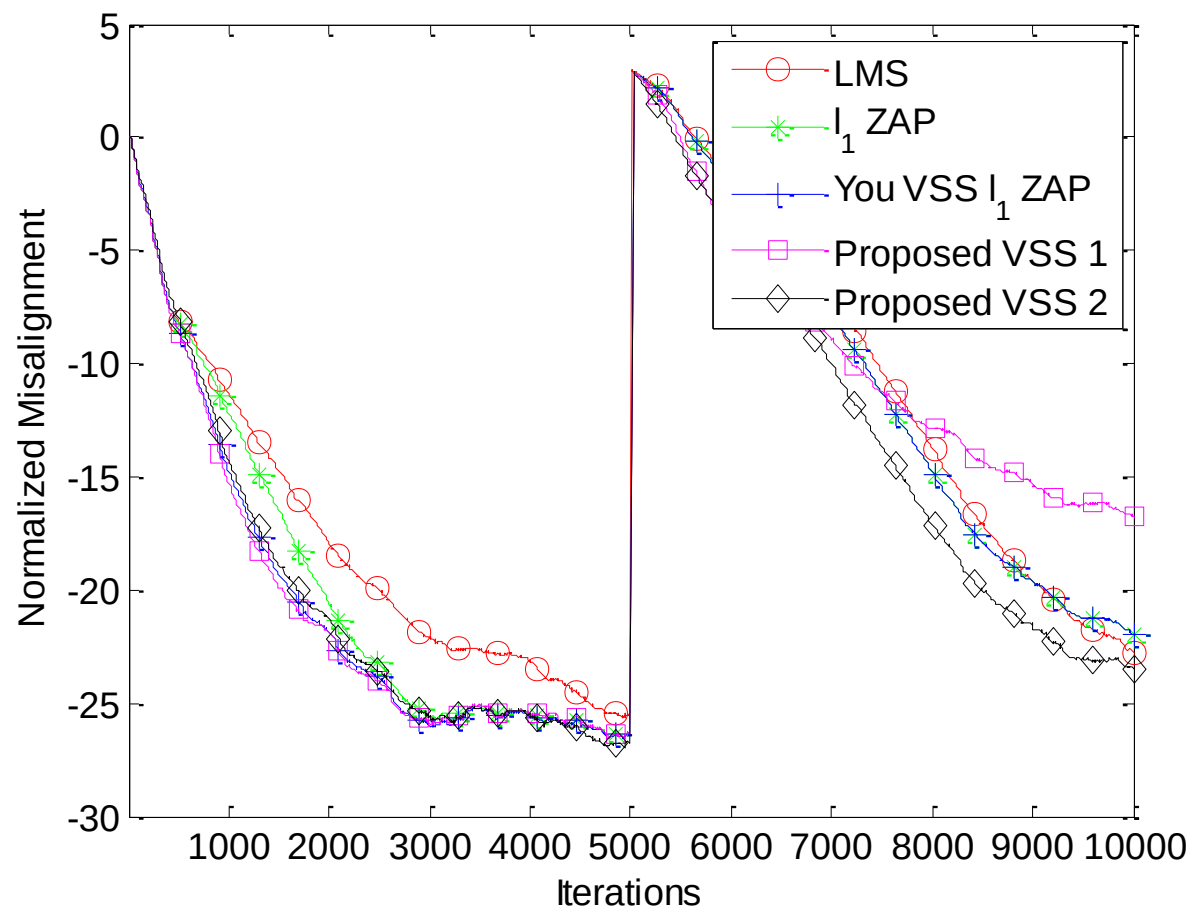

**Fig.6a** Comparison of normalized misalignment for $l_1$ norm constraint ZAP under dispersive system.

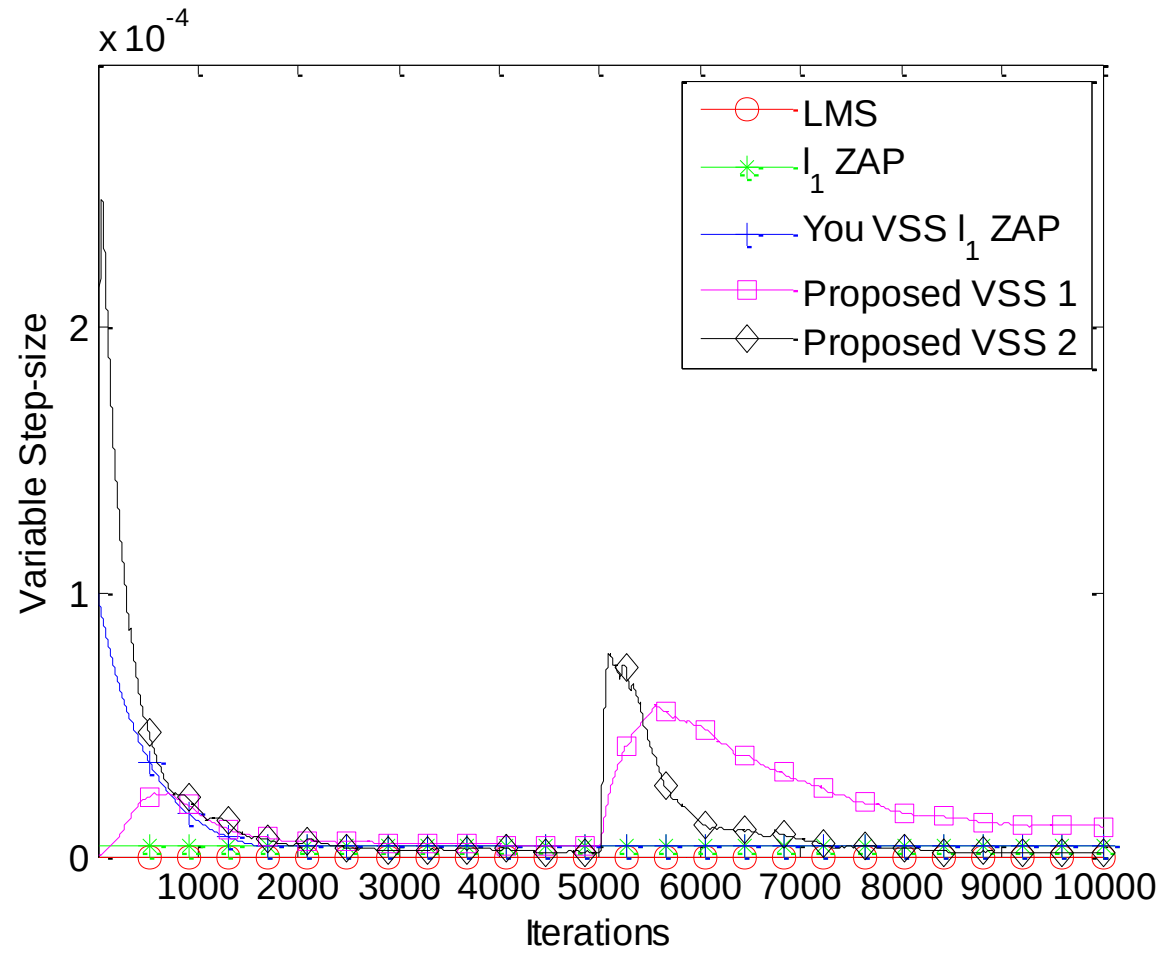

**Fig.6b** Comparison of variable step-size for $l_1$ norm constraint ZAP under dispersive system.

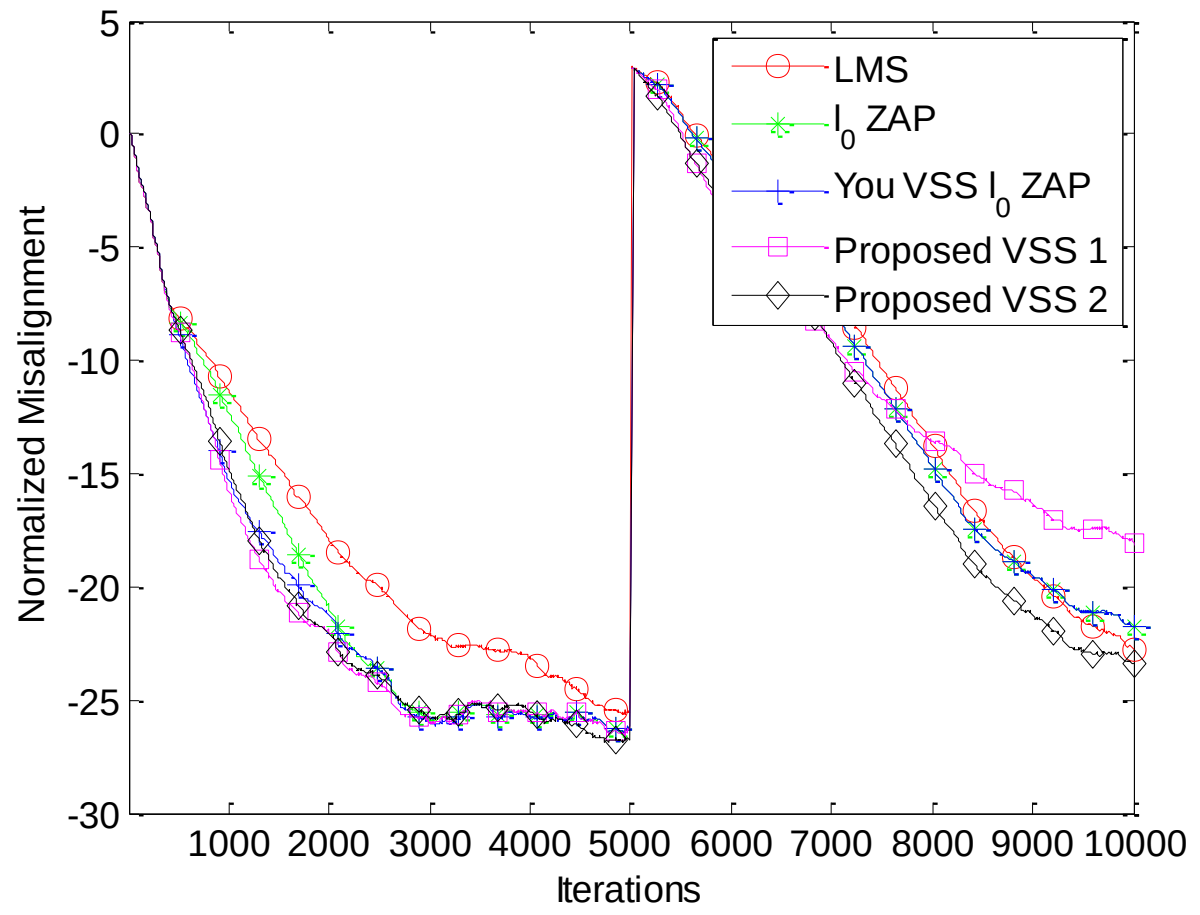

**Fig.7a** Comparison of normalized misalignment for $l_0$ norm constraint ZAP under dispersive system.

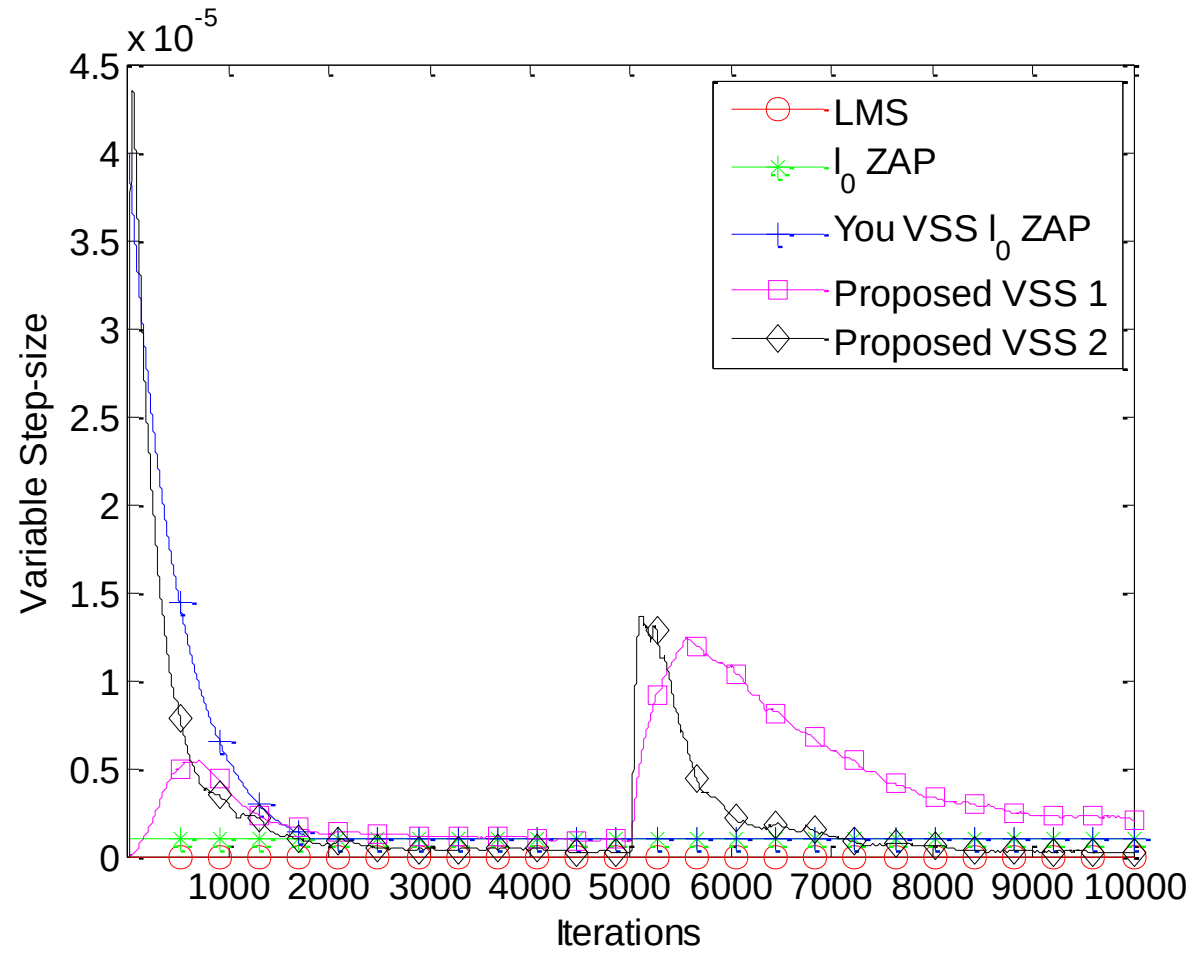

**Fig.7b** Comparison of variable step-size for $l_0$ norm constraint ZAP under dispersive system.

## 5. CONCLUSION

A new variable step-size scheme for the zero-point attraction projection algorithm was proposed in this paper, which is based on the estimation of sparseness gradient. Simulation results demonstrate that, for sparse system identification, the proposed VSS ZAP could provide both faster convergence rate and better tracking ability than previous VSS algorithms. Meanwhile, it could remove the impact of ZAP term for dispersive impulse response, which is preferable.